\documentclass[10pt,journal,letter]{IEEEtran}
\usepackage{graphics}
\usepackage{bm}
\usepackage{amsmath}
\usepackage{epstopdf}
\usepackage{graphicx}
\usepackage{amsmath}
\usepackage{amsfonts}
\usepackage{color}
\usepackage[]{algorithm2e}
 \usepackage{tikz}

\let\MYoriglatexcaption\caption
\renewcommand{\caption}[2][\relax]{\MYoriglatexcaption[#2]{#2}}

\renewcommand{\eqref}[1]{(\ref{#1})}

\begin{document}

\title{Random Access in C-RAN for User Activity Detection with Limited-Capacity Fronthaul}

\author{\begin{tabular}{c}
Zoran~Utkovski,~\IEEEmembership{Member,~IEEE,}
Osvaldo~Simeone,~\IEEEmembership{Fellow,~IEEE,}
Tamara~Dimitrova,\\  
and Petar~Popovski,~\IEEEmembership{Fellow,~IEEE}
\end{tabular}
\thanks{Z.~Utkovski is with Faculty of Computer Science, University Goce Del\v{c}ev \v{S}tip, with Department of Signal Theory and Communications, University Carlos III of Madrid, and with Laboratory for Complex Systems and Networks, Macedonian
Academy of Sciences and Arts. O.~Simeone is with the Center for Wireless Information Processing (CWIP), New Jersey Institute of Technology. T.~Dimitrova is with Laboratory for Complex Systems and Networks, Macedonian
Academy of Sciences and Arts. P.~Popovski is with the Department of Electronic Systems, Aalborg University.} }

\maketitle
\vspace{-10pt}
\begin{abstract}
Cloud-Radio Access Network (C-RAN) is characterized by a hierarchical structure in which the baseband processing functionalities of remote radio heads (RRHs) are implemented by means of cloud computing at a Central Unit (CU). A key limitation of C-RANs is given by the capacity constraints of the fronthaul links connecting RRHs to the CU. In this letter, the impact of this architectural constraint is investigated for the fundamental functions of random access and active User Equipment (UE) identification in the presence of a potentially massive number of UEs. In particular, the standard C-RAN approach based on quantize-and-forward and centralized detection is compared to a scheme based on an alternative CU-RRH functional split that enables local detection. Both techniques leverage Bayesian sparse detection. Numerical results illustrate the relative merits of the two schemes as a function of the system parameters.
\end{abstract}
\vspace{-5mm}
\section{Introduction}
In a Cloud-Radio Access Network (C-RAN) the baseband processing functionality is implemented at a centralized cloud processor or Central Unit (CU) on behalf of multiple distributed Remote Radio Heads (RRHs). This is made possible by the fronthaul links that connect the RRHs to the CU, with possible media that include fiber optic cables, DSL last-mile links or wireless mmwave channels. The capacity limitations of the fronthaul links, along with the associated latency, are understood to offer the most significant challenge to the implementation of C-RANs \cite{checko2014cloud}.

A fundamental network function is random access, which is carried out by user equipments (UEs) when first accessing the system. Random access is attracting renewed interest due to the expected increase in the number of UEs in 5G networks, with particular reference to massive access in Internet-of-Things applications (see, e.g., \cite{laya2014random}). One of the main goals of the random access procedure is for the network to identify the set of active UEs in order to enable resource allocation. In the context of C-RANs, random access for initial access has a rather novel aspect, as there is no single radio access point to which the terminal is associated. 

In this letter, we study \textit{user activity detection} (UAD) 
for a C-RAN architecture with the aim of investigating solutions that address the mentioned fronthaul capacity limitations. We assume that the UEs employ non-orthogonal sequences so as to accommodate a potentially massive number of UEs, e.g., machine-type devices, and we do not assume any \textit{a priori} knowledge of the instantaneous small-scale fading channel realizations. Under the further assumption that the number of active UEs is significantly smaller than the total number of UEs, the signal received at the RRH is \textit{sparse} with respect to the set of the UE signatures. As a result, the UAD problem becomes one of sparse signal recovery 
(see, e.g., \cite{fletcher2009off, Calderbank, Wunder14} and references therein). A sparsity-based algorithm for UAD in C-RANs was recently proposed in \cite{Lau15} using a Bayesian formulation under the assumption of ideal, i.e., infinite-capacity, fronthaul links. 
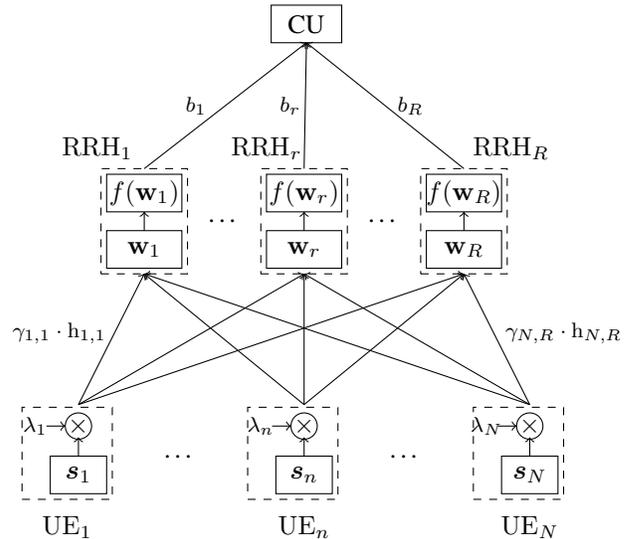
\begin{figure}
\centering
\begin{tikzpicture}
\draw (0.5-0.75/2,0.25) rectangle (1.25-0.75/2,0.75);
\node at (0.5,0.5) {$\bm{s}_1$};
\node at (1.85,0.75){\dots};

\draw (3.5-0.75/2,0.25) rectangle (4.25-0.75/2,0.75);
\node at (3.5,0.5) {$\bm{s}_n$};
\node at (4.85,0.75){\dots};

\draw (6.5-0.75/2,0.25) rectangle (7.25-0.75/2,0.75);
\node at (6.5,0.5) {$\bm{s}_N$};

\draw (0.5,1.15) circle [radius=0.175];
\node at (0.5,1.15) {$\times$};
\draw [->] (0.5-0.175-0.25,1.35-0.2) -- (0.5-0.175,1.35-0.2);
\node at (0.5-0.325-0.25,1.35-0.2){\footnotesize{$\mathrm{\lambda}_1$}};
\draw [dashed] (0.5-0.325-0.25-0.175,0.25-0.075) rectangle (1.25-0.75/2+0.075,1.35-0.2+0.175+0.075);

\draw (3.5,1.15) circle [radius=0.175];
\node at (3.5,1.15) {$\times$};
\draw [->] (3.5-0.175-0.25,1.35-0.2) -- (3.5-0.175,1.35-0.2);
\node at (3.5-0.325-0.25,1.35-0.2){\footnotesize{$\mathrm{\lambda}_n $}};
\draw [dashed] (3.5-0.325-0.25-0.175,0.25-0.075) rectangle (4.25-0.75/2+0.075,1.35-0.2+0.175+0.075);

\draw (6.5,1.15) circle [radius=0.175];
\node at (6.5,1.15) {$\times$};
\draw [->] (6.5-0.175-0.25,1.35-0.2) -- (6.5-0.175,1.35-0.2);
\node at (6.5-0.325-0.25,1.35-0.2){\footnotesize{$\mathrm{\lambda}_N $}};
\draw [dashed] (6.5-0.325-0.25-0.175,0.25-0.075) rectangle (7.25-0.75/2+0.075,1.35-0.2+0.175+0.075);

\draw [->] (0.5,0.75) -- (0.5,1);
\draw [->] (3.5,0.75) -- (3.5,1);
\draw [->] (6.5,0.75) -- (6.5,1); 

\draw (1-0.125,3.25) rectangle (1.75+0.125,3.75);
\node at (2.75/2,3.5) {$\mathbf{w}_1$};
\node at (2.75/4+3.5/2,3.5+0.375) {\dots};

\draw (1-0.125,4) rectangle (1.875,4.5);
\node at (2.75/2,4.25) {$f(\mathbf{w}_1)$};
\draw [->] (2.75/2,3.75) -- (2.75/2,4);

\draw [dashed] (1-0.2,3.25-0.075) rectangle (1.875+0.075,4.5+0.075);
\draw [dashed] (3.5-0.75/2-0.125-0.075,3.25-0.075) rectangle (4.25-0.75/2+0.125+0.075,4.5+0.075);
\draw [dashed] (5.25-0.125-0.075,3.25-0.075) rectangle (6+0.125+0.075,4.5+0.075);

\draw (3.5-0.75/2-0.125,3.25) rectangle (4.25-0.75/2+0.125,3.75);
\node at (3.5,3.5) {$\mathbf{w}_r$};
\node at (3.5/2+4.25/2+2.75/4,3.5+0.375) {\dots};

\draw (3.5-0.75/2-0.125,4) rectangle (4.25-0.75/2+0.125,4.5);
\node at (3.5,4.25) {$f(\mathbf{w}_r)$};
\draw [->] (3.5,3.75) -- (3.5,4);

\draw (5.25-0.125,3.25) rectangle (6+0.125,3.75);
\node at (3.25+1+2.75/2,3.5) {$\mathbf{w}_R$};

\draw (5.25-0.125,4) rectangle (6+0.125,4.5);
\node at (3.25+1+2.75/2,4.25) {$f(\mathbf{w}_R)$};
\draw [->] (3.25+1+2.75/2,3.75) -- (3.25+1+2.75/2,4);

\draw (3.5-0.875/2,6.25) rectangle (4.25+0.125-0.75/2,6.75);
\node at (3.5/2-0.875/4+4.25/2+0.125/2-0.75/4,6.5) {CU};


\draw [->] (0.5,1.35+0.085) -- (3.5,3.25-0.075);
\draw [->] (3.5,1.35+0.085) -- (3.5,3.25-0.075);
\draw [->] (6.5,1.35+0.085) -- (3.5,3.25-0.075);

\draw [->] (0.5,1.35+0.085) -- (1+0.75/2,3.25-0.075);
\draw [->] (3.5,1.35+0.085) -- (1+0.75/2,3.25-0.075);
\draw [->] (6.5,1.35+0.085) -- (1+0.75/2,3.25-0.075);

\draw [->] (0.5,1.35+0.085) --  (5.25+0.75/2,3.25-0.075);
\draw [->] (3.5,1.35+0.085) -- (5.25+0.75/2,3.25-0.075);
\draw [->] (6.5,1.35+0.085) -- (5.25+0.75/2,3.25-0.075);

\draw [->] (3.5,3.25-0.075+1.4) -- (3.5/2-0.875/4+4.25/2+0.125/2-0.75/4,6.25);
\draw [->] (1+0.75/2,3.25-0.075+1.4) -- (3.5/2-0.875/4+4.25/2+0.125/2-0.75/4,6.25);
\draw [->] (5.25+0.75/2,3.25-0.075+1.4) -- (3.5/2-0.875/4+4.25/2+0.125/2-0.75/4,6.25);

\node at (0.25,1.35/2+0.175/2+3.25/2){\footnotesize{$\gamma_{1,1}\cdot \mathrm{h}_{1,1}$}};
\node at (6.95,1.35/2+0.175/2+3.25/2){\footnotesize{$\gamma_{N,R}\cdot \mathrm{h}_{N,R}$}};

\node at (3.5/2+3.5/4-0.875/8+4.25/4+0.125/4-0.75/8-0.15-1.3, 3.25/2-0.075/2+1.4/2+6.25/2){\footnotesize{$b_{1}$}};

\node at (3.5/2+3.5/4-0.875/8+4.25/4+0.125/4-0.75/8-0.2, 3.25/2-0.075/2+1.4/2+6.25/2){\footnotesize{$b_{r}$}};

\node at (3.5/2+3.5/4-0.875/8+4.25/4+0.125/4-0.75/8-0.2+1.6, 3.25/2-0.075/2+1.4/2+6.25/2){\footnotesize{$b_{R}$}};

\node at (0.75,4.85){$\mathrm{RRH}_1$};
\node at (3,4.85){$\mathrm{RRH}_r$};
\node at (6.25,4.85){$\mathrm{RRH}_R$};

\node at (0.35,-0.2){$\mathrm{UE}_1$};
\node at (3.5,-0.2){$\mathrm{UE}_n$};
\node at (6.5,-0.2){$\mathrm{UE}_N$};
\end{tikzpicture}
\caption{C-RAN architecture performing random access: $R$ RRHs serve $N$ randomly activated UEs which use sequences of length $M$. Each RRH $r$ computes a function of the received signal, $f(\mathbf{w}_r)$, where $f(\cdot)$ depends on the type of processing performed at the RRH and is limited to $b_r$ bits per sample. The output is forwarded to the CU for joint processing over fronthaul links $\mathrm{RRH}_r-\mathrm{CU}$ of capacity $b_r$ bits per sample.}
\label{fig:System_Model}
\end{figure}

With the aim of investigating the impact of fronthaul capacity limitations, in this letter we first study  
the standard C-RAN implementation, whereby the RRHs quantize the received samples for transmission on the fronthaul links to the CU, which performs centralized baseband processing for UAD (see, e.g., \cite{checko2014cloud}). To the best of our knowledge, the impact of quantization on UAD has not been studied to date. Furthermore, we also consider a baseline scheme that adopts an alternative CU-RRH functional split, in the sense of, e.g., \cite{dotsch2013quantitative}, in which part of the baseband processing is carried out at the RRHs. In particular, each of the RRHs performs local UAD, and forwards quantized soft information associated with the local decision in the form of log-likelihood ratios (LLRs) to the CU. The two schemes, which are referred to as \textit{Quantize-and-Forward} (QF) and \textit{Detect-and-Forward} (DtF), are compared via numerical results in terms of the trade-off between the \textit{fraction of correctly detected active UEs} and the \textit{fraction of incorrectly detected inactive UEs}.


\textit{Notation:} Uppercase/lowercase boldface letters denote matrices/vectors. Random quantities are represented with standard fonts, while italic is used for deterministic quantities. The superscript $\mathrm{H}$ stands for
Hermitian transposition and $\otimes$ stands for Kronecker matrix multiplication. 
$\mathcal{N}(\mu,\sigma^2)$ and $\mathcal{CN}(\mu,\sigma^2)$ denote real and circularly symmetric, respectively, Gaussian random variable with expectation $\mu$
and variance $\sigma^2$. The notation $\left[\mathbf{X}_1,\ldots,\mathbf{X}_N\right]$ for matrices $\mathbf{X}_1,\ldots,\mathbf{X}_N$ of suitable sizes represents a matrix that stacks $\mathbf{X}_1,\ldots,\mathbf{X}_N$ vertically, while $\left[\mathbf{X}_1; \ldots; \mathbf{X}_N\right]$ stacks $\mathbf{X}_1,\ldots,\mathbf{X}_N$ horizontally. 

\vspace{-6pt}
\section{System Model}
\label{sec:System_Model}
We consider a slotted random access system with $N$ UEs. The user activity (random)  
variable $\mathrm{\lambda}_n\in\{0,1\}$ equals $1$ if
UE $n$ is active in the given block, which happens with probability $p$, and $0$ otherwise. When active, UE $n$ transmits over $M$, in general, complex  symbols of the time-frequency grid the identification signature 
\begin{equation}
\bm{s}_n
=\left[
  s_{n,1},\:\cdots\:,s_{n,M}\right].
\end{equation}   
The UE signatures are subject to the energy constraint $\mathrm{E}\left[\|\mathbf{s}_n\|^2\right]=E_s$. We assume that the time-frequency grids of different users are aligned, which requires time synchronization\footnote{Our model may be extended to include frame asynchronicity (with symbol-level  synchronization intact) by using cyclic-extended signature waveforms based, for example, on Gabor frames or Kerdock codes \cite{Calderbank}.}. Moreover, focusing on a scenario with a potentially massive number $N$ of UEs, the signatures are assumed to be nonorthogonal. Assuming a block-fading model with coherence time-frequency span no smaller than that occupied by the signatures' transmission, the signal  
$\mathbf{w}_r=\left[
  \mathrm{w}_1,\:\cdots\:,\mathrm{w}_M\right]$ received at RRH $r$, $r=1,\ldots R$, reads
\begin{equation}
\mathbf{w}_r=\sum_{n=1}^N\mathrm{\lambda}_n\mathrm{\gamma}_{n,r}\mathrm{h}_{n,r}\bm{s}_n+\mathbf{v}_r,
\label{eq:system_model}
\end{equation}
where $\mathrm{\gamma}_{n,r}\in \mathbb{R}$ and $\mathrm{h}_{n,r}\in \mathbb{C}$ are the large- and small-scale fading coefficients, respectively, for the link between the $n$-th UE and the $r$-th RRH, and $\mathbf{v}_r$ is an additive noise vector. We further assume that the coefficients $\mathrm{h}_{n,r}$ are independent identically distributed (i.i.d.) $\mathcal{CN}(0,1)$ and the elements of the noise vector $\mathbf{v}_r$ are i.i.d. $\mathcal{CN}(0,\sigma_v^2)$. \
The large scale fading coefficients $\mathrm{\gamma}_{n,r}$ are assumed to be known to the RRHs and to the CU as in, e.g., \cite{Lau15}, in contrast to the unknown small scale fading coefficients $\mathrm{h}_{n,r}$. We assume that the UEs, RRHs and the CU know the small-scale fading statistics and the probability of UE activation $p$. 

The fronthaul capacity limitations are expressed in terms of the number of bits per received complex sample, $b_r$ available for transmission on the fronthaul link between RRH $r$ and the CU. The system model is illustrated in Fig. \ref{fig:System_Model}. To elaborate further, we rewrite the received signal (\ref{eq:system_model}) as
\begin{align}
\mathbf{w}_r&=\bm{S}\bm{\Gamma}_r\mathbf{\Lambda}\mathbf{h}_r+\mathbf{v}_r,
\label{eq:system_model_2}
\end{align}
where the columns of $\bm{S}\in \mathbb{C}^{M\times N}$ represent the UEs' signatures; $\bm{\Gamma}_r\in \mathbb{R}^{N\times N}$ is a diagonal matrix with $\gamma_{n,r}$, $n=1,\ldots,N$, on the main diagonal; $\mathbf{h}_r\in \mathbb{R}^{N\times 1}$ is a vector of the small-scale fading coefficients; and $\mathbf{\Lambda}$ is a diagonal matrix  
with vector $\bm{\lambda}=\left[
\mathrm{\lambda_1},\:\cdots\:,\mathrm{\lambda_N}\right]$ on the main diagonal. We can further simplify (\ref{eq:system_model_2}) as 
\begin{equation}
\mathbf{w}_r=\bm{A}_r\mathbf{x}_r+\mathbf{v}_r,
\label{eq:System_Model_Bernoulli-Gauss}
\end{equation}
with the definitions $\bm{A}_r = \bm{S}\bm{\Gamma}_r$ and $\mathbf{x}_r = \mathbf{\Lambda}\mathbf{h}_r$. Note that the columns of $\bm{A}_r$ are the signatures of the UEs scaled with the corresponding large scale fading coefficients of the links between the UEs and $r$-th RRH, which are assumed to be known to the RRH, while $\mathbf{x}_r$ depends on the unknown user activity and small-scale fading variables. 

The model (\ref{eq:System_Model_Bernoulli-Gauss}) can be interpreted within a Bayesian formulation of the sparse detection problem (see, e.g., \cite{foucart2013mathematical}). In fact, the unknown $\mathbf{x}_r\in \mathbb{C}^{N\times 1}$ is a Bernoulli-Gaussian random vector, with entries $\mathrm{x}_{n,r}$, $n=1,\ldots,N$ being equal to the channel $h_{n,r}\sim\mathcal{CN}(0,1)$ with probability $p$, accounting for the case of an active UE $n$ ($\lambda_n=1$), or else equal to zero with probability $1-p$ when UE $n$ is inactive ($\lambda_n=0$). When $p$ is small, and in the absence of fronthaul capacity limitations, UAD hence translates into estimating the support of a sparse i.i.d. Bernoulli-Gaussian vector from a Multiple Measurement Vector (MMV) model (see, e.g., \cite{foucart2013mathematical}), as investigated in \cite{Lau15}. 
\vspace{-7pt}
\section{Fronthaul and UAD Processing} 
In this section, we discuss two baseline schemes that account for fronthaul capacity limitations, namely QF and DtF.
\vspace{-20pt} 
\subsection{Quantize-and-forward (QF)}
\label{sec:QF}   
With QF, each RRH $r$ quantizes the measurement $\mathbf{w}_r$ in (\ref{eq:System_Model_Bernoulli-Gauss}) with $b_r$ bits per sample, and forwards the quantized samples to the CU.  The signal received by the CU on the $r$-th fronthaul can hence be written as 
\begin{align}
\mathbf{y}_r=\mathcal{Q}_r(\mathbf{w}_r)=\mathcal{Q}_r(\bm{A}_r\mathbf{x}_r+\mathbf{v}_r),
\label{eq:System_Model_Quantization}
\end{align}
where
$\mathcal{Q}_r$ is a quantization function, applied element-wise to the entries of the argument vector, with resolution $b_r$ bits. We assume here that the function $\mathcal{Q}_r$ amounts to two scalar uniform quantizers with $2^{b_r/2}$ levels applied to the real and imaginary parts of the entries of $\mathbf{w}_r$. The dynamic range is selected so as to capture three standard deviations for both positive and negative values of each real component. 

The overall signal  $ \mathbf{y}=\left[\mathbf{y}_1,\ldots,\mathbf{y}_R\right]$ retrieved by the CU from the fronthaul links can then be expressed in a compact fashion by defining $\bm{\Gamma}^{(n)}$ as the diagonal matrix with the vector $\left[\gamma_{n,1},\dots,\gamma_{n,R}\right]$ of long-term fading coefficients on the main diagonal along with the unknown vector $\mathbf{x}^{(n)}=\left[x_{n,1},\ldots,x_{n,R}\right]$. In particular, we can write
\begin{align}
\mathbf{y}=\mathcal{Q}\left(\mathbf{w}\right)
=\mathcal{Q}\left(\bm{A}\mathbf{x}+\mathbf{v}\right),
\label{eq:System_Model_QF}
\end{align}
where we have 
$
\bm{A}=\left[\bm{\Gamma}^{(1)}\otimes \mathbf{s}_1; \ldots; \bm{\Gamma}^{(N)}\otimes \mathbf{s}_N\right]$, $\mathbf{x}=\left[\mathbf{x}^{(1)},\ldots,\mathbf{x}^{(N)}\right]$,  $\mathbf{w}=\left[\mathbf{w}_1,\ldots,\mathbf{w}_R\right]$ and $\mathcal{Q}$ is to be understood as being the same as $\mathcal{Q}_r$ whenever it is applied to a component of $\mathbf{w}$ coming from $\mathbf{w}_r$. 

The CU performs UAD based on the received signal (\ref{eq:System_Model_QF}), by implementing a sparsity-based reconstruction algorithm, e.g. in the spirit of compressive sensing (CS), that aims at estimating the support of the Bernoulli-Gaussian vector $\mathbf{x}$ from the linear mixture $\mathbf{z}+\mathbf{v}$, with $\mathbf{z}=\bm{A}\mathbf{x}$, as observed after the application of the sample-by-sample non-linearity $\mathcal{Q}$. In particular, the unknown vector $\mathbf{x}$ in (\ref{eq:System_Model_QF}) is characterized by \emph{group sparsity}, since each subvector $\mathbf{x}^{(n)}$ equals an all-zero vector if UE $n$ is not active, i.e., if $\lambda_n=0$, and is generally non-zero when $\lambda_n=1$. For the purpose of signal reconstruction, here we adopt the Hybrid Generalized Approximate Message Passing (H-GAMP) method developed in \cite{Rangan12}, \cite{RanganHGAMP}, which extends over the Generalized AMP scheme (GAMP) \cite{Rangan11} and accommodates both nonlinear measurements (i.e. quantization) and group sparsity. Details of the GAMP implementation for de-quantization in compressive sensing may be found in \cite{Kamilov}.
    
H-GAMP, as GAMP,  is based on a quadratic approximation of the sum-product message passing scheme and operates by exchanging messages on the factor graph that describes the joint distribution $p_{\bm{\lambda},\mathbf{x},\mathbf{y}}(\bm{\lambda},\bm{x},\bm{y})$. More precisely, since the H-GAMP algorithm operates on real-valued variables, we first redefine the signal model (\ref{eq:System_Model_QF}) as follows: (\emph{i}) each entry of the vectors $\mathbf{x}$ and $\mathbf{y}$ is substituted by two real entries corresponding to its real and imaginary parts; (\emph{ii}) each of the entries $a_{ij}$ of matrix $\bm{A}$ is substituted by the submatrix \begin{small}$\left(\begin{array}{cc}\mathrm{Re}(a_{ij})&-\mathrm{Im}(a_{ij})\\\mathrm{Im}(a_{ij})&\mathrm{Re}(a_{ij})\end{array}\right)$\end{small}. Note that each subvector $\mathbf{x}^{(n)}$ is now of size $2R$, instead of $R$, but the group sparsity properties of the vector $\mathbf{x}$ are preserved. 

Denoting by $\mathbf{x}$ and $\mathbf{y}$ the real-valued vectors introduced above, the joint distribution of $\bm{\lambda}$, $\mathbf{x}$ and $\mathbf{y}$ over which the H-GAMP algorithm operates factors as 
\begin{small}
\begin{equation}
p_{\bm{\lambda},\mathbf{x},\mathbf{y}}(\bm{\lambda},\bm{x},\bm{y}) = \prod_{n=1}^{N} p_{\mathrm{\lambda}}(\lambda_n) \prod_{j=1}^{2RN} p_{\mathrm{x}|\lambda}(x_j|\lambda_{\xi(j)}) \prod_{i=1}^{2RM} p_{\mathrm{y}|\mathrm{z}}(y_i | z_i),
\label{eq:posterior}
\end{equation}
\end{small}where $\xi(j)$ denotes the index of the UE that corresponds to entry $x_j$; $\mathrm{z}_i=\bm{a}_i^{\mathrm{T}}\mathbf{x}$, with $\bm{a}_i$ being the $i$-th row of $\bm{A}$; and $\mathrm{v}_i\sim\mathcal{N}(0,\sigma_v^2/2)$. In (\ref{eq:posterior}), we have $p_{\mathrm{\lambda}}(\lambda)=p^{\lambda}(1-p)^{1-\lambda}$; $p_{\mathrm{x}|\lambda}(x_{j}|\lambda_{\xi(j)})$ amounts to the $\mathcal{N}(0,1/2)$ probability density function if $\lambda_{\xi(j)}=1$, and to a Kronecker delta function centred at $x_j=0$ if $\lambda_{\xi(j)}=0$; and 
\begin{equation}
p_{\mathrm{y}|\mathrm{z}}(y_i|z_i)=\int_{\mathcal{Q}^{-1}(\mathrm{y}_i)}\phi(u;z_i,\sigma_v^2/2)\:\mathrm{d}u,
\label{eq:Output}
\end{equation}
where $\mathcal{Q}^{-1}$ is the inverse of the component of the quantization function $\mathcal{Q}$ that applies to the entry $y_i$ and $\phi(\cdot;\mu,\sigma^2)$ denotes the Gaussian probability density function with mean $\mu$ and variance $\sigma^2$. Note that the factorization (\ref{eq:posterior}) uses the fact that, for a given UE activity pattern $\bm{\lambda}$, the small-scale channel coefficients in $\mathbf{x}$ are independent. 

The H-GAMP algorithm, which is detailed in \cite{Rangan12} and \cite{RanganHGAMP}, can be directly applied to (\ref{eq:posterior}) to output an approximation of the posterior distributions $p_{\mathrm{\lambda}|\mathbf{y}}(\lambda_n|\bm{y})$ for all UE $n=1,...,N$. From these probabilities, the log-likelihood ratio (LLR) $l_n$ associated with the belief that UE $n$ is active is computed as $l_n=\log \frac{p_{\mathrm{\lambda}|\mathbf{y}}(\lambda_n=1|\bm{y})}{p_{\mathrm{\lambda}|\mathbf{y}}(\lambda_n=0|\bm{y})}$. Based on the LLR $l_n$, the CU estimates the user activity variable as $\hat{\mathrm{\lambda}}_n=1$ if $l_n\geq l_{th}$ for some threshold $l_{th}$ and $\hat{\mathrm{\lambda}}_n=0$ otherwise. The details of the H-GAMP implementation are provided in Appendix  \ref{sec:Appendix}. 

\textit{Remark 1:} In the presence of a packetized fronthaul transmission, e.g., via Ethernet, instead of a rate constraint on the fronthaul link, it is relevant to consider a constraint on the overall number of bits $B=Mb$ that the RRH can communicate to the CU. In this case, it is possible to trade the signature length, say $M'$, with the number of bits per complex sample, say $b'$, under the constraint $B=M'b'$. In the single measurement vector (SMV) setting, this problem has been studied in the framework of recovery of sparse signals from $1$-bit measurements in \cite{Jacques13},\cite{Plan14}, and for quantized measurements with multiple quantization levels in \cite{Rosasco13}. 

\vspace{-10pt}
\subsection{Detect-and-forward (DtF)}
With DtF, each of the RRHs performs a \emph{local estimate} of the UEs' activity pattern $\{\lambda_n\}$ and then forwards quantized soft information on these estimates to the CU over the capacity limited fronthaul links. Specifically, each RRH $r$ calculates 
$l_{r,n}=\log \frac{p_{\mathrm{\lambda}|\mathbf{w}_r}(\lambda_n=1|\bm{w}_r)}{p_{\mathrm{\lambda}|\mathbf{w}_r}(\lambda_n=0|\bm{w}_r)}$, 
which is the LLR associated with the belief of UE $n$ being active based on the observation $\mathbf{w}_r$ in (\ref{eq:System_Model_Bernoulli-Gauss}). This can be done by means of the H-GAMP following the same approach discussed above. Each RRH then quantizes each LLR as 
$\widetilde{l}_{r,n}=\mathcal{Q}_{r}(l_{r,n})$, 
where $\mathcal{Q}_{r}$ is a scalar quantization function applied to the (real-valued) argument. Given the fronthaul rate $b_r$, the quantizer $\mathcal{Q}_{r}$ has a number of levels equal to $2^{Mb_r/N}$ since there are $N$ LLRs to quantize with a total of $Mb_r$ bits. The dynamic range is selected based on preliminary Monte Carlo simulations to capture a 95\% confidence interval. 
 
UAD detection is finally performed at the CU by summing the LLRs obtained from all the RRHs. Specifically, for the UE $n$, the CU computes $\widetilde{l}_{n}=\sum_{r=1}^R\widetilde{l}_{r,n}$ and then applies a threshold test on the resulting LLR $\widetilde{l}_{n}$. We observe that this test is optimal, in the case of unquantized LLRs, in case the observations of the RRHs are conditionally i.i.d. given the transmitted signatures, while this is, in general, not the case here due to different large-scale fading coefficients $\gamma_{n,r}$. Generalized rules that capture asymmetries in the quality of the observations at the RRHs can be devised but they will not be further investigated here. 

\textit{Remark 2:} In this letter, we consider scalar quantization for both QF and DtF. It should be mentioned that improved performance could be obtained by leveraging compression techniques in lieu of scalar quantization. For instance, for DtF, the RRH could first perform an estimate of the Bernoulli vector associated with the UE activity pattern and then compress it losslessly using the knowledge of the probability $p$ (see \cite{Weidmann12} for related discussion on the compression of sparse sources).

\section{Numerical Results}
\label{sec:Results}
In this section, we perform numerical analysis with the aim to capture the relative merits of the standard C-RAN implementation with centralized baseband processing, and the alternative RRH-CU functional split in which part of the baseband processing is performed at the RRHs, under fronthaul capacity constraints. In particular,  
we compare the UAD performance of QF and DtF as a function of the key system parameters given by the number of the UEs $N$, the probability of UE activation $p$, the signature length $M$, the number of RRHs $R$, and the fronthaul capacity $b_r$.  

For simplicity, in the following we assume a dense network with large-scale fading coefficients $\gamma_{n,r}=1$ for all pairs of UEs and RRHs. The average signal-to-noise ratio (per system user) is defined as $
\rho\doteq E_s/(M\sigma_v^2)$. In the following, we assume the UE signature vectors to be random, with i.i.d. circularly-symmetric complex Gaussian entries, for which the convergence of approximate message passing schemes has been studied rigorously \cite{Donoho}, \cite{Rangan11, Rangan12}. We note that the described schemes are not tied to the specific choice of the signature sequences, and other constructions, e.g., based on Gabor frames or Kerdock/Reed-Mueller codes, may be used. In that case, however, the convergence of H-GAMP and approximate message passing in general has to be addressed accordingly (see \cite{Caltagirone14} for related discussion).
\begin{figure}[h!]
\centering
\includegraphics[width=92mm]{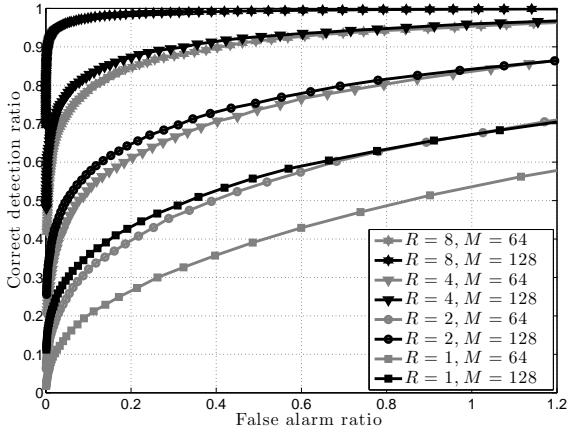}
\caption{Correct detection ratio vs. false alarm ratio for QF for different values of RRH $R$ ($N=256$, $p=48/256$, $b_r=4$ for $M=128$, and $b_r=8$ for $M=64$, 
and SNR per user $\rho=-10.81\,\mathrm{dB}$).} 
\label{fig:Figure_1}
\end{figure}

Fig. \ref{fig:Figure_1} investigates the effect of the number of RRHs $R$ on the UAD performance of the QF scheme. Specifically, we plot the \textit{correct detection ratio} versus the \textit{false alarm ratio}, where the former is the ratio of the number of correctly detected users over the number of active users, and the latter is the ratio of the inactive UEs that are detected as being active over the number of active users. Note that these performance metrics reflect the classical trade-off in hypothesis testing between the two types of probability of error, accounting for incorrect detection and false alarm events. Furthermore, the curves are obtained by varying the thresholds in the tests introduced in the text above. The fronthaul rate is fixed to $b_r=4$ bits per (complex) symbol for $M=128$, so that the total number of fronthaul bits per packet is $Mb_r=512$ (see Remark 1). We also consider a shorter signature of $M=64$, in which case we set $b_r=8$ in order to keep the same total number of fronthaul bits. 

From Fig.  \ref{fig:Figure_1}, we first observe the expected trade-off between correct detection ratio and false alarm ratio. More interestingly, we note the sharp improvement in the performance with the increase of the number of RRHs $R$, ranging here $1$ to $8$ -- an effect similar to the one observed with the \textit{rank-aware} CS reconstruction algorithms in \cite{Eldar}. Finally, we note that trading the signature length $M$ for a larger quantization depth $b_r$ deteriorates the performance, demonstrating the advantages of using a larger signature over using a more refined quantization. 

\begin{figure}[t]
\centering
\includegraphics[width=92mm]{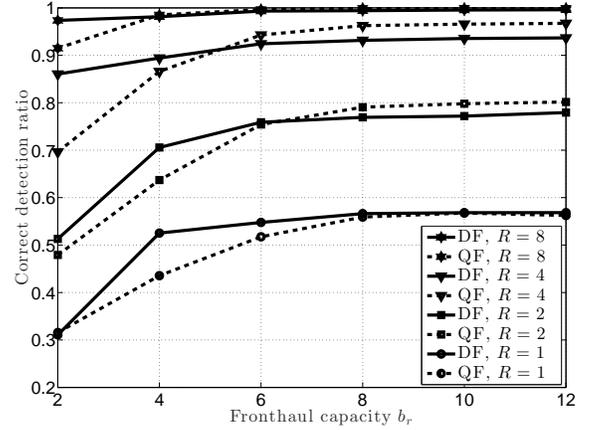}
\caption{Correct detection ratio of QF and DtF as function of the fronthaul capacity $b_r$ for a false alarm ratio equal to $0.2$ ($N=256$, $p=48/256$, SNR per user $\rho=-10.81 \, \mathrm{dB}$).} 
\label{fig:Figure_2}
\end{figure}

Fig. \ref{fig:Figure_2} aims at investigating the impact of fronthaul limitations on the performance of QF and DtF. 
The comparison between the two schemes is performed here under a fixed average false alarm ratio of $0.2$ (see Fig. \ref{fig:Figure_1} for an illustration). The signature length is $M=128$ and we consider different values for the number $R$ of RRHs. We observe that DtF outperforms QF for stringent fronthaul capacity constraints. This is due to the fact that the performance of QF under a small fronthaul bit budget is hampered by coarseness of the fronthaul quantization (see \cite{Goyal08} for related discussion). Instead, DtF benefits from the local UAD processing done at the RRHs to reduce the amount of information that needs to be transmitted to the CU (see also \cite{Weidmann12}). 
In the complementary regime in which the fronthaul capacity is sufficiently large, QF outperforms DtF. In fact, when the quantized signals are sufficiently accurate, QF benefits from the joint processing capabilities of the CU to perform UAD on the signals received by the RRHs. This is unlike DtF, in which local UAD processing prevents the CU to have direct access to the measurements of the RRHs. Finally, it is observed that the performance of UAD saturates at (relatively) moderate values of $b_r$, demonstrating that in this regime the performance is limited by the signature length $M$ rather than by the fronthaul capacity constraints. 
\section{Conclusions}
We investigated the impact of fronthaul capacity limitations in a C-RAN architecture on the functions of random access and active UE identification in the presence of a potentially massive number UEs. In particular, we studied the performance of two baseline algorithmic solutions leveraging Bayesian sparse detection: a standard C-RAN approach based on quantize-and-forward (QF) and an alternative scheme based on detect-and-forward (DtF). Numerical results illustrate the relative merits of the two schemes. While here we have concentrated on the function of user activity detection, our framework could also be extended to integrate data transmission, by assigning a subset of sequences to serve as codewords for each of the system users. Future interesting work includes the analysis of the impact of more sophisticated compression techniques for fronthaul transfer on the performance of random access. 
\appendices
\section{Details on H-GAMP Algorithm}
\label{sec:Appendix}
\subsection{Overview}
Under the graphical model (\ref{eq:posterior}), Appendix C in \cite{RanganHGAMP} shows that the sum-product version of Hybrid-GAMP algorithm reduces to the GAMP procedure in \cite{Rangan11} run in a parallel with updates of the sparsity levels. Specifically, each iteration $t$ has two stages. The first half of the iteration, labeled as the "basic GAMP update", is identical to the standard updates from the basic GAMP algorithm \cite{Rangan11}, treating the components $\mathrm{x}_j$ as independent with sparsity level $\widehat{\rho}_j^t$. The second half of the  iteration, labeled as the "sparsity update", updates the sparsity levels $\widehat{\rho}_j^t$ based on the estimates from the basic GAMP half of the algorithm. The quantity $\widehat{\rho}_j^t$ is an estimate for the probability that the component $\mathrm{x}_j$ belongs to an active group, i.e. that the UE with index $\xi(j)$ is active, as we are dealing with nonoverlapping groups. In the following, we will use $G_n$ to denote the group (set) of $2R$ indices of the vector $\mathbf{x}$ associated with UE $n$, $n=1,\ldots, N$. The details of the algorithm may be obtained directly from \cite{Rangan12}, \cite{RanganHGAMP} (main H-GAMP references) and \cite{Kamilov} (which includes details about the GAMP implementation in the context of de-quantization in compressive sensing), but are summarized below for reference.
\subsection{GAMP Update}
Due to the assumed independence of the components $x_j$ during the GAMP update in iteration $t$, the GAMP algorithm operates on the factor graph described by \begin{equation} p_{\mathbf{x}|\mathbf{y}}(\bm{x}|\bm{y})\propto \prod_{j=1}^{2RN} p_{\mathrm{x}}(x_j)\prod_{i=1}^{2RM} p_{\mathrm{y}|\mathrm{z}}(y_i|z_i), \label{eq:posterior_GAMP} \end{equation} with the aim of estimating $\mathbf{x}$ from the observation vector $\mathbf{y}$. In (\ref{eq:posterior_GAMP}) the symbol $\propto$ denotes identity after normalization to unity. For the prior of $\mathbf{x}$, which is updated in each iteration $t$, we have that $\mathrm{x}_j\sim \mathcal{N}(0,1/2)$ with probability $\widehat{\rho}_j^t$, and $\mathrm{x}_j=0$ with probability $1-\widehat{\rho}_j^t$.

By reserving the indices $j, k \in \{1,\ldots, 2RN\}$ for the variable nodes, and $i, l \in \{1,\ldots, 2RM\}$ for the factor nodes, in a belief propagation (BP) setting, the following messages are passed  along the edges of the graph (\ref{eq:posterior_GAMP}) \begin{eqnarray} \mu_{i\leftarrow j}^t(x_j)&\propto& p_{\mathrm{x}}(x_j)\prod_{l\neq i} \mu_{l\rightarrow j}^t(x_j),\\ \mu_{i\rightarrow j}^t(x_j)&\propto& \int p_{\mathrm{y}|\mathrm{z}} (y_i|z_i)\prod_{k\neq j}\mu_{i\leftarrow k}^{t-1}(x_j)\:\mathrm{d}\,\mathrm{x}_{|j}. \label{eq:BP_GAMP} \end{eqnarray}
In (\ref{eq:BP_GAMP}) the integration is over all elements of $\mathbf{x}$ except $x_j$.  

From ({\ref{eq:BP_GAMP}}) the approximate marginal distribution is computed as \begin{equation} \widehat{p}_{\mathrm{x}_j|\mathbf{y}}(x_j|\bm{y})\propto p_{\mathrm{x}}(x_j)\prod_{i=1}^{2RM} \mu_{i\rightarrow j}^t(x_j). \end{equation} Finally, the component $\widehat{x}_j^t$ of the estimate $\widehat{\bm{x}}^t$ is computed as \begin{equation} \widehat{x}_j^t=\int_{\mathbb{R}}x \widehat{p}_{x_j|\mathbf{y}}(x|\bm{y})\,\mathrm{d} x. \end{equation}
GAMP relies on a Gaussian approximation to overcome the complexity of the BP message passing scheme. Details of the implementation of the GAMP algorithm for a generalized linear mixing problem with quantized outputs are presented in [12]. The only difference here is that the distribution of the components of $\mathbf{x}$ is updated in each iteration $t$ during the sparsity level update. Following [12], here we simply  summarize the GAMP equations, with the discussed sparsity level adaptation in mind.

\subsubsection{Initialization}
We start by setting/evaluating  \begin{eqnarray} \widehat{\mathbf{x}}^0 &=&\mathrm{E}[\mathbf{x}],\\ \mathbf{v}_{x}^0 &=&\mathrm{var}[\mathbf{x}],\\ \widehat{\mathbf{s}}^0&=&\mathbf{0}, \end{eqnarray} where the expectation and the variance are with respect to the (Bernoulli-Gaussian) prior $p_{\mathbf{x}}$. The initial values of the sparsity levels are set to $\rho_j^0=p$, $j=1,\ldots, 2RN$, where $p$ is the probability of a UE being active.
\subsubsection{Factor Update}
In the factor update we first compute the linear step \begin{eqnarray} \widehat{\mathbf{p}}^t&=&\bm{A}\widehat{\mathbf{x}}^{t-1}-\mathbf{v}_{p}^t\bullet\widehat{\mathbf{s}}^{t-1},\\\mathbf{v}_{p}^t &=&(\bm{A}\bullet\bm{A})\mathbf{v}_{x}^{t-1}, \end{eqnarray} where $\bullet$ denotes the Hadamard product (component-wise multiplication). Then, we  evaluate the nonlinear step \begin{eqnarray} \widehat{\mathbf{s}}^t&=&\mathrm{E_1}\left(\mathbf{y},\widehat{\mathbf{p}}^t,\mathbf{v}_{p}^t+\frac{\sigma_v^2}{2}\mathbf{e};\mathcal{Q}\right),\\ \mathbf{v}_{\mathbf{s}}^t&=&\mathrm{V_1}\left(\mathbf{y},\widehat{\mathbf{p}}^t,\mathbf{v}_{p}^t+\frac{\sigma_v^2}{2}\mathbf{e};\mathcal{Q}\right), \end{eqnarray} where $\mathbf{e}$ is the all-ones vector.

The scalar functions $\mathrm{E_1}$ and $\mathrm{V_1}$, applied  over the components of $\mathbf{y}$, are defined as \begin{eqnarray} \mathrm{E_1}\left(\mathrm{y},\widehat{p},v_{p};\mathcal{Q}\right)&=&\frac{1}{v_p}\left(\mathrm{E}\left[\mathrm{z}|\mathrm{z}\in \mathcal{Q}^{-1}(y)\right]-\widehat{p}\right),\\ \mathrm{V_1}\left(\mathrm{y},\widehat{p},v_{p};\mathcal{Q}\right)&=&\frac{1}{v_p}\left(1-\frac{\mathrm{var}\left[\mathrm{z}|\mathrm{z}\in \mathcal{Q}^{-1}(y)\right]}{v_p}\right)\label{eq:V_1}, \end{eqnarray} where the expectation and the variance are evaluated with respect to $\mathrm{z}\sim \mathcal{N}(\widehat{p},v_{p})$.
\subsubsection{Variable Update}
In the variable update we first compute the linear step \begin{eqnarray} \widehat{\mathbf{r}}^t&=&\widehat{\mathbf{x}}^{t-1}+\mathbf{v}_{r}^t\bullet(\bm{A}^{\mathrm{T}}\widehat{\mathbf{s}}^t),\\ \mathbf{v}_{r}^t&=&\left((\bm{A}\bullet\bm{A})^\mathrm{T}\mathbf{v}_{s}^t\right)_{\circ}^{-1}, \end{eqnarray} where $(\cdot)_\circ^{-1}$ denotes component-wise exponentiation. Then, we evaluate the nonlinear step \begin{eqnarray} \widehat{\mathbf{x}}^t&=&\mathrm{E_2}\left(\widehat{\mathbf{r}}^t,\mathbf{v}_{r}^t; p_{\mathbf{x}}^t\right),\\ \mathbf{v}_{x}^t&=&\mathrm{V}_2\left(\widehat{\mathbf{r}}^t,\mathbf{v}_{r}^t; p_{\mathbf{x}}^t\right), \end{eqnarray} where the superscript $t$ in $p_{\mathrm{x}}^t$ is used to denote that the distribution of $\mathbf{x}$ is updated in each iteration $t$. The scalar functions $\mathrm{E}_2$ and $\mathrm{V}_2$ are applied component-wise and are given by \begin{eqnarray} \mathrm{E}_2\left(\widehat{r},v_{r}; p_{\mathrm{x}}\right)&=&\mathrm{E}[\mathrm{x}|\widehat{r}],\\ \mathrm{V}_2\left(\widehat{r},v_r;p_{\mathrm{x}}\right)&=&\mathrm{var}[\mathrm{x}|\widehat{r}].\label{eq:V_2} \end{eqnarray} The expected value and the variance are evaluated with respect to  \begin{equation} p_{\mathrm{x}|\widehat{\mathrm{r}}}(\cdot|\widehat{r})\propto \phi\left(\cdot;\widehat{r},v_{r}\right)p_{\mathrm{x}}(\cdot). \end{equation} The above equations may be interpreted as a denoising process operating on the scalar variables $\mathrm{x}$ and $\widehat{\mathrm{r}}$, where: $\mathrm{x}\sim\mathcal{N}(0,1/2)$ with probability $\widehat{\rho}$, and $\mathrm{x}=0$ with probability $1-\widehat{\rho}$; $\widehat{\mathrm{r}}$ is an AWGN-corrupted version of $\mathrm{x}$, namely \begin{equation} \widehat{\mathrm{r}}=\mathrm{x}+\mathrm{w}, \:\: \mathrm{w}\sim  \mathcal{N}(0,v_r). \label{eq:vr} \end{equation}
\subsection{Sparsity Level Update}
Based on the output of the GAMP part of the iteration, the sparsity update stage updates the quantity $\widehat{\rho}_j^t$ which, we recall, represents an estimate for the probability that the UE with index $\xi(j)$ is active. 

We start by defining the message \begin{equation} l_{j\rightarrow n}^t=\log \left(\frac{p_{\widehat{\mathrm{r}}}(\widehat{r}_j^t;\rho_j=1)}{p_{\widehat{\mathrm{r}}}(\widehat{r}_j^t;\rho_j=0)}\right) \label{eq:LLR_j_to_n}. \end{equation} 
As we are dealing with nonoverlapping groups, the message (\ref{eq:LLR_j_to_n}) is the ratio of likelihood of the output $\widehat{r}$ given that $x_j$ belongs to an active group (i.e. $\lambda_{\xi(j)}=1$) to the likelihood given that $x_j$ does not belong to an active group (i.e. $\lambda_{\xi(j)}=0$).   From (\ref{eq:vr}) we have \begin{equation} l_{j\rightarrow n}^t=\log \frac{\phi\left(\widehat{r}_j^t;0,0.5+v_r\right)}{\phi\left(\widehat{r}_j^t;0,v_r\right)}, \label{eq:LLR_j_to_n_expression} \end{equation} where $\phi(\cdot;\mu,\sigma^2)$ denotes the pdf of a Gaussian scalar random variable with mean $\mu$ and  variance $\sigma^2$. 

Defined in this way, $l_{j\rightarrow n}^t$ may be understood as a (local) estimate of the log-likelihood ratio \begin{equation} l_n=\log \frac{p_{\mathrm{\lambda}|\mathbf{y}}(\lambda_n=1|\bm{y})}{p_{\mathrm{\lambda}|\mathbf{y}}(\lambda_n=0|\bm{y})}.  \end{equation} The estimate is updated by "collecting" the messages corresponding to all indices from the group $G_n$ (except $j$)     \begin{equation} l_{j\leftarrow n}^t=\log\left(\frac{p}{1-p}\right)+\sum_{k\in G_n, \,k\neq j} l_{k \rightarrow n}^t.  \end{equation} Finally, the procedure returns the estimate \begin{equation} \widehat{\rho}_{j}^{t+1}=1-\frac{1}{1+\mathrm{exp}\left(l_{j\leftarrow n}^{t}\right)}, \end{equation} which is used in the next iteration of the GAMP update part of the algorithm.
\bibliographystyle{IEEEtran}
\bibliography{Bibliography}

\end{document}